\documentclass{article}                            
\usepackage{amsmath,amssymb,citesort,enumerate,eucal,supertabular}

\def\bn{\bigskip\noindent}

\def\sn{\smallskip\noindent}

\title{Transit of Luyten 726-8 within 1 ly from Epsilon Eridani}
\author{Igor Yu. Potemine\footnote{Universit\'e Paul Sabatier, Institut de 
Math\'ematiques de Toulouse (IMT), 118 route de Narbonne, F-31062 
Toulouse Cedex 9, France, igor.potemine@math.univ-toulouse.fr}}
\date{}

\begin{document}
\maketitle

\abstract{This is one of results from our program of massive simulations
of close encounters for all nearby stars. Epsilon Eridani is an extremely 
interesting star having one confirmed planet and multiple asteroid and 
debris belts. It should have a quite massive Oort cloud as well. Deltorn 
et al. searched for past Nemesis encounters of $\varepsilon$ Eri. In this 
paper we show that, according to current astrometric data, an other famous 
nearby star Luyten 726-8AB (=BL/UV Ceti) will pass at $\lesssim$ 0.93 ly
from Epsilon Eridani in $\approx$ 31.5 kyr. So, it will probably pierce 
through the outer part of the hypothetical Oort cloud of $\varepsilon$ Eri. 
BL/UV Ceti has only about 20 percent of the solar mass. Nevertheless, it 
could influence directly some long-period comets of Epsilon Eridani. The 
duration of mutual transit of two star systems within 1 ly from each other is
$\gtrsim$ 4.6 kyr. Our simulations show that stellar encounters within 1 ly 
might be more frequent than previously thought. It could explain Proxima's 
peculiar trajectory with respect to $\alpha$ Cen AB or even Sedna's 
trajectory in the solar system.

\section{Introduction}

Epsilon Eridani is a relatively young K-star having one confirmed planet, 
two asteroid belts at $\sim3$ and $\sim20$ AU resp.~as well as an analog 
of Kuiper belt starting at 35 AU from the star. It should also have the 
outer cloud of long-period comets more massive than the Oort cloud in our 
solar system. The litterature on $\varepsilon$ Eri is very extensive, 
\emph{cf.}~1159 references currently available in {\footnotesize SIMBAD} 
database for additional details.   

\smallskip BL/UV Ceti was discovered by Luyten in 1948-49 as a star with a 
very high proper motion. UV Ceti is actually an archetype of flare stars 
famous for its extreme changes of brightness. Two components are at few 
AU from each other on highly elliptical orbits. The total mass of the 
system is about $0.2M_{\odot}$. Luyten 726-8 is also the closest currently 
known neighbour of Epsilon Eridani (\emph{cf.}~{\footnotesize SIMBAD} 
database for an extensive list of references).

\smallskip Deltorn and Kalas \cite{D} searched for past close encounters of 
$\varepsilon$ Eri. They have discovered that Epsilon Eridani had recently
a close encounter with Kapteyn's star at the mutual distance of $\approx$ 
1 pc. So, as far as I know, it is somewhat unexpected that it will have a 
much closer encounter just with its currently nearest neighbour!

\section{Distance and time of the closest encounter}
\label{sect:dist}

The linear approximation gives very accurate results within $\pm 100$ kyr.
Let us denote Epsilon Eridani by $X$ and BL/UV Ceti by $Y$. Using 
{\footnotesize HIP2} data for $\varepsilon$ Eri as well as 
{\footnotesize RECONS} parallax and {\footnotesize SIMBAD} data for BL/UV 
Ceti, we get the following relative position of $Y$ with respect to $X$ :
\begin{equation}
\vec{r}_{XY}^{}=(4.5962,2.0840,-0.6559)\textrm{ ly}
\end{equation}
in rectangular galactic coordinates and Julian light-years. Thus, the current 
distance $d_{XY}=\Vert\vec{r}_{XY}^{}\Vert$ between Epsilon Eridani and 
Luyten 726-8 is $\approx$ 5.09 ly. Radial velocities of stars $X$ and $Y$ are 
$v_{r,X}^{}=16.3\pm0.1$ ({\footnotesize PCRV}, \cite{G}) and 
$v_{r,Y}^{}=21.9\pm0.1$ (Malaroda et al., \cite{M}). We will also use 
an older measurement $v_{r,Y}^{}= 29\pm2$ from {\footnotesize GCRV}.

\sn\tablecaption{{\footnotesize HIP2} and {\footnotesize PCRV} data for 
$\varepsilon$ Eri; {\footnotesize RECONS} parallax and radial velocities 
from Malaroda+ and {\footnotesize GCRV} for BL/UV Ceti. }

\tablehead{\hline}\tabletail{\hline}
\begin{supertabular}{|c|c|c|c|c|c|c|}
Id &$\pi_X^{}$ &$v_t$ &$v_r$ &U &V &W\\
&(mas) &(km/s) &(km/s) &(km/s) &(km/s) &(km/s)\\

\hline $\varepsilon$ Eri &310.74 &14.87 &+16.3 &-3.54 &+7.09 &-20.59\\
\hline {\footnotesize BL/UV Cet} &373.70 &42.42 &+21.9 &-41.83 &-19.40 
&-12.36\\
& & &+29 &-43.57 &-19.27 &-19.24\\
\end{supertabular}

\bn The transverse velocity $v_t^{}$ is easily calculated from the parallax 
and total proper motion (\emph{cf.}~\cite{P}, sect.~2). So, we get the 
following relative space velocities:
\begin{equation}
\vec{\mu}_{XY}^{} = (-38.29,-26.49,+8.23)\textrm{ or }(-40.03,-26.36,+1.35)
\end{equation}
The time (with respect to J2000.0 epoch) and distance of the closest 
encounter are given by the formulas:
\begin{align}
t_{XY}^{\textrm{min}} &=-\frac{\langle \vec{\mu}_{XY}^{},\vec{r}_{XY}^{}
\rangle\times c}{\Vert\vec{\mu}_{XY}\Vert^2}\quad\textrm{yr},\\
d_{XY}^{\textrm{min}} &=\sqrt{d_{XY}^{\,2} -
\frac{\langle \vec{\mu}_{XY}^{},\vec{r}_{XY}^{}\rangle^2}
{\Vert\vec{\mu}_{XY}^{}\Vert^2}}
\quad\textrm{ly,}
\end{align}
where $\langle\cdot,\cdot\rangle$ is the standard scalar product of vectors
and $c$ is the speed of light in km/s. Using these formulas, we obtain:
\begin{equation}
d_{XY}^{\textrm{min}} = \mathbf{0.93}\pm0.01\textrm{ ly and }
t_{XY}^{\textrm{min}} = \mathbf{31.5}\pm0.2\textrm{ kyr}
\end{equation}
for the closest encounter of $\varepsilon$ Eri with Luyten 726-8. Actually, 
intermediate radial velocities (in the range from +22 to +29 km/s)
for BL/UV Ceti give even better results. For example, if $v_{r,Y}^{}=+25$ km/s
then $d_{XY}^{\textrm{min}} \approx \mathbf{0.86}$ ly. 

\smallskip Finally, the duration of mutual transit 
$\Delta t_{XY}^{(1\,\textrm{ly})}$ of two star systems within 1 ly from 
each other is given by the formula:
\begin{equation}
\Delta t_{XY}^{(1\,\textrm{ly})}=2t_{XY}^{\textrm{min}}
\times\sqrt{\frac{1-(d_{XY}^{\textrm{min}})^2}
{d_{XY}^{\,2}-(d_{XY}^{\textrm{min}})^2}}\quad\textrm{yr}
\end{equation}
So, we get $\Delta t_{XY}^{(1\,\textrm{ly})}\approx \textbf{4.6}$ kyr from the 
main data and $\Delta t_{XY}^{(1\,\textrm{ly})}\approx \textbf{6.4}$ kyr if 
$v_{r,Y}^{}=+25$ km/s.
 
\section{Discussion}

This and other examples from our simulations show that stellar encounters 
within 1 ly might be more frequent than previously thought. It could explain 
peculiar trajectories of some stars, brown dwarves, planets and minor 
celestial bodies. For instance, Proxima's trajectory with respect to 
Alpha Centauri AB might be the result of one or several close encounters. 
Such encounters could even exlain Sedna's trajectory in the solar system!

\section{Acknowledgements}

This research has made use of the SIMBAD database, operated at CDS, 
Strasbourg, France.

\nocite{*}
\bibliographystyle{plain}
\bibliography{epseri}

\end{document}